\documentclass[]{tJOT2e}

\usepackage{times} 
\usepackage{graphicx} 
\usepackage{dcolumn}
\usepackage{bm}
\usepackage{amssymb}
\usepackage{amsmath}
\usepackage{wasysym}
\usepackage{color}
\usepackage{hyperref}
\usepackage[normalem]{ulem}
\usepackage{float}
\usepackage{flushend}
\usepackage{multirow}
\usepackage{cancel}
\usepackage[dvipsnames]{xcolor}
\hypersetup{
colorlinks,
citecolor=blue,
filecolor=blue,
linkcolor=blue,
urlcolor=blue}
\newcommand\redsout{\bgroup\markoverwith{\textcolor{red}{\rule[0.5ex]{2pt}{1pt}}}\ULon}

\begin{document}

\markboth{Abhishek Kumar, Mahendra K. Verma, and Jai Sukhatme}{Journal of Turbulence}

\articletype{}

\title{Phenomenology of two-dimensional stably stratified turbulence under large-scale forcing}
\vskip 0.3 in

\author{Abhishek Kumar$^{\rm a \ast}$, \thanks{$^\ast$Corresponding author. Email: abhkr@iitk.ac.in
\vspace{6pt}} Mahendra K. Verma$^{\rm a}$, and 
\vspace{6pt} Jai Sukhatmae$^{\rm b}$\\\vspace{6pt}  $^{\rm a}${\em{Department of Physics, Indian Institute of Technology, Kanpur 208016, India}}\\$^{\rm b}${\em{Centre for Atmospheric and Oceanic Sciences, Indian Institute of Science, Bangalore 560012, India.}}\\$^{\rm c}${\em{Divecha Centre for Climate Change, Indian Institute of Science, Bangalore 560012, India.}} }

\maketitle

\begin{abstract}
In this paper we characterize the scaling of energy spectra, and the interscale transfer of energy and enstrophy, for strongly, moderately
and weakly
stably stratified two-dimensional (2D) turbulence under large-scale
random forcing. In the strongly stratified case, a large-scale verically sheared horizontal flow (VSHF) co-exists with small scale
turbulence. The VSHF consists of internal gravity waves and the turbulent flow has a
kinetic energy (KE) spectrum that follows an approximate $k^{-3}$ scaling with zero KE flux and a robust positive enstrophy flux. The
spectrum of the turbulent potential
energy (PE) also approximately follows a $k^{-3}$ power-law and its flux is directed to small scales.
For moderate stratification, there is no VSHF and the KE of the turbulent flow exhibits Bolgiano-Obukhov scaling that
transitions from a shallow
$k^{-11/5}$ form at large
scales, to
a steeper approximate $k^{-3}$ scaling at small scales. The entire range of scales shows a strong forward enstrophy flux, and interestingly,
large (small) scales show an inverse (forward) KE flux. The PE flux in this regime is directed
to small scales, and the PE spectrum is characterized by an approximate $k^{-1.64}$ scaling.
Finally, for weak stratification,
KE is transferred upscale and its spectrum closely follows
a $k^{-2.5}$ scaling, while PE exhibits a forward transfer and its spectrum shows an approximate $k^{-1.6}$ power-law. For all stratification strengths,
the total energy always flows from large to small scales and almost all the spectral indicies are well explained by accounting for
the scale dependent nature of the corresponding flux.
\end{abstract}


\section{Introduction}

Stable stratification with rotation is an important feature of geophysical flows \cite{Vallis:Book}.  The strength of stratification is usually measured by the non-dimensional parameter called Froude number ($\mathrm{Fr}$), which is defined as the ratio of the time scale of gravity waves and the nonlinear time scale.  Strong stratification has $\mathrm{Fr} \ll  1$, while weak stratification has  $\mathrm{Fr}  \ge 1$~\cite{LR-rev}.  Researchers have studied such flows  with all or some of the features. In this paper we restrict ourselves to two-dimensional stable stratification \cite{Lilly,Hopfinger} that allows us to explore a wide range of $\mathrm{Fr}$ and characterise the interscale transfer of energy and enstrophy, and  the energy spectra in strongly, moderately, and weakly stratified scenarios.  

Apart from a reduction in dimensionality, the 2D stratified system differs from the more traditional three-dimensional (3D) equations in that it lacks a vortical mode. Indeed, the decomposition of the 3D system into vortical and wave modes \cite{Leith,LR-rev,EM} has proved useful in studying stratified \cite{LR-1991,LMD,WB1,WB2,LB-2007} and rotating-stratified \cite{Bartello,SW:JFM,KM-grl,SS-gafd,VDL,MDHP} turbulence. The absence of a vortical mode implies that the 2D stratified system only supports nonlinear wave-wave mode interactions \cite{CF-1983,FB-1983} 
(interestingly, 3D analogs that only support wave interactions have also been considered previously \cite{LY,RSS2}). 
In a decaying setting, the initial value problem concerning the fate of a standing wave in 2D has been studied experimentally \cite{BS-1996} 
and numerically \cite{BSS1,BSS2}. The regime was that of strong stratification, and not only where the waves observed to break, 
this process was accompanied by a forward energy transfer
due to nonlocal parametric
subharmonic instability \cite{BSS1}. Further, at long times after breaking, the turbulence generated was 
characterised by a $k_{\parallel}^{-3}$ scaling 
(i.e., parallel to
the direction of ambient stratification) \cite{BSS2}. Other decaying simulations, that focussed on the formation and distortion of fronts
from an initially smooth profile, noted a self-similarity in the probability density function of the vorticity field as well as more of an isotropic
$k^{-5/3}$ kinetic energy (KE) spectrum, though at early stages in the evolution of the system \cite{Sukhatme:POF2007}. 

With respect to the forced problem, the case of random small scale forcing has been well studied. Specifically, at moderate stratification,
the 2D system developed a robust 
vertically sheared horizontal flow (VSHF; \cite{SW:JFM}) accompanied by an inverse transfer of KE and a $k^{-5/3}$ scaling \cite{Smith:CM2001}. 
For weak 
stratification, a novel flux loop mechanism involving the upscale transfer of KE (with $k^{-5/3}$ scaling) and the downscale transfer of 
potential energy (PE), also with $k^{-5/3}$ scaling,
was seen to result in a stationary state \cite{Boffetta:EPL2011}. In fact, moisture driven strongly stratitifed flows in 2D have also
been seen to exhibit a upscale KE transfer with a $k^{-5/3}$ scaling \cite{Sukhatme:POF2012}.
Interestingly, large-scale forcing in the form of a temperature gradient has been examined experimentally. Specifically, using a soap film,
Zhang {\em et al.}~\cite{Zhang:PRL2005} reported scaling of density fluctuations at low frequencies with exponents $-7/5$ (Bolgiano scaling~\cite{Bolgiano:JGR1959, Obukhov:DANS1959})
and $-1$ (Batchelor scaling~\cite{Batchelor:JFM1959a}) for moderate and strong temperature gradients, respectively. Seychelles {\em et al.}~\cite{Seychelles:PRL2008}
noted a similar scaling, and the development of isolated coherent vortices on a curved 2D soap bubble.

In the present work we look at the relatively unexamined case of random forcing at large scales.
The flows are simulated using a pseudospectral code Tarang~\cite{Verma:Pramana2013}
for $\mathrm{Fr}$ ranging from $0.16$ to $1.1$.
For $\mathrm{Fr} = 0.16$, i.e., strong stratification, a VSHF (identified as internal gravity waves) emerges at large scales
and co-exists with small scale turbulence. The
turbulent flow is characterized by a forward enstrophy cascade, zero KE flux and a KE spectrum that scales approximately as $k^{-3}$. The PE
spectrum also follows an approximate $k^{-3}$ power-law with a scale dependent flux of the form $k^{-2}$. At moderate stratification,
there is no VSHF, and the KE spectrum shows a modified form of Bolgiano-Obukhov ~\cite{Bolgiano:JGR1959,Obukhov:DANS1959} scaling for
2D flows --- approximately $k^{-11/5}$ at large scales and $k^{-3}$ at small scales.
The KE flux also changes character with scale, and exhibits an inverse (forward) transfer at large (small) scales.
The PE spectrum follows an approximate $k^{-1.64}$ scaling and its flux is weakly scale dependent. Finally, for $\mathrm{Fr} = 1.1$, i.e., weak
stratification, the KE flux is upscale for most scales and its spectrum is characterized by a $-2.5$ exponent. The PE flux
continues to be downscale and its spectrum obeys an approximate $k^{-1.6}$ scaling. All exponents observed are well explained by
taking into account the variable nature of the corresponding flux. Exceptions are the PE spectra for moderate and weak
stratification whose scaling is little steeper than expected.

The outline of the paper is as follows: In Sec.~\ref{sec:gov_eqn}, we describe the equations governing stably-stratified (SS) flows and the associated parameters. In Sec.~\ref{sec:sim_method}, we discuss the numerical details of our simulations. In the subsequent three subsections, we detail various kinds of flows observed for strongly SS flows in Sec.~\ref{sec:gravity_wave}, moderately SS flows in Sec.~\ref{sec:bolgiano}, and weakly SS flows in Sec.~\ref{sec:weak_stratification}. Finally, we conclude in Sec.~\ref{sec:conclusion} with a summary and discussion of our results.

\section{Governing Equations}
\label{sec:gov_eqn}
We employ the following set of equations to describe 2D stably stratified flows \cite{Sutherland}:
\begin{eqnarray}
\frac{\partial {\bf u}}{\partial t} + ({\bf u} \cdot \nabla) {\bf u} & = & -\frac{1}{\rho_0} \nabla \sigma+ \alpha g \theta \hat{z} + \nu \nabla^2 {\bf u} + {\bf f}_u, \label{eq:u_dim} \\
\frac{\partial \theta}{\partial t} + (\bf u \cdot \nabla) \theta & = & - \frac{d\bar{T}}{dz} u_z + \kappa \nabla^2 \theta, \label{eq:th_dim} \\
\nabla \cdot \bf u & = & 0 \label{eq:inc_dim}, 
\end{eqnarray}
where $\bf u$ is the 2D velocity field, $\theta$ and $\sigma$ are respectively the temperature and pressure fluctuations from the conduction state, 
$\hat{z}$ is the buoyancy direction while $\hat{x}$ is the horizontal direction, ${\bf f}_u$ is the external force field, $g$ is the acceleration 
due to gravity, and $\rho_0$, $\alpha$, $\nu$, and $\kappa$ are the fluid's mean density, thermal expansion coefficient, kinematic viscosity, 
and thermal diffusivity, respectively. In the above description, we make the Boussinesq approximation under which the density 
variation of the fluid is neglected except for the buoyancy term. Also, $d\bar{T}/dz > 0$ due to stable stratification.  Note that we use temperature  rather than density as a variable.  We can convert Eqs.~(\ref{eq:u_dim}-\ref{eq:inc_dim})  in terms of density $\rho'$ using the following relations:
\begin{equation}
\frac{\rho'}{\rho_0} = -\alpha \theta; ~~~~ \frac{d \bar{\rho}}{dz} = -\rho_0 \alpha \frac{d\bar{T}}{dz}.
\end{equation}

The linearised version of the  Eqs.~(\ref{eq:u_dim}-\ref{eq:inc_dim}) yields internal gravity waves for which the velocity and density 
fluctuate along $\hat{z}$ with \text{Brunt V\"{a}is\"{a}l\"{a}} frequency,
\begin{equation}
N = \sqrt{\alpha g  \frac{d\bar{T}}{dz}}.
 \label{eq:Frequency}
\end{equation}
The linearised equations also yield a dispersion relation for the internal gravity waves as
\begin{equation}
\Omega = N\frac{k_x}{k},
 \label{eq:dispersion}
\end{equation}
where  $k = \sqrt{k_x^2 + k_z^2}$.   In addition to $N$, the important nondimensional variables used for describing SS flows are,
\begin{eqnarray}
\text{Prandtl number } \mathrm{Pr} & = & \frac{\nu}{\kappa}, \label{eq:pr} \\
\text{Rayleigh number } \mathrm{Ra} & = & \frac{d^4 \alpha g}{\nu\kappa} \left |\frac{d\bar{T}}{dz} \right |,  \label{eq:ra} \\
\text{Reynolds number } \mathrm{Re} & = &\frac{u_{rms}d}{\nu}, \label{eq:Re} \\
\text{Richardson number }  \mathrm{Ri} & = &\frac{\alpha g d^2}{u_{rms}^2}\left |\frac{d\bar{T}}{dz} \right |, \label{eq:Ri} \\
\text{Froude number }  \mathrm{Fr} & =  &\frac{u_{rms}}{d N}, \label{eq:Fr}
\end{eqnarray}
where $u_{rms}$ is the rms velocity of flow, which is computed as a volume average. 
Note that the Richardson number is the ratio of the buoyancy and the nonlinearity $(\bf u \cdot \nabla) \bf u$, 
and is related to the Froude number as   $\mathrm{Ri} = 1/\mathrm{Fr}^2$. 

In the limiting case of $\nu = \kappa =0$ and ${\bf f}_u = 0$, the total energy 
\begin{equation}
 E =   \frac{1}{2} \int  \left[ u^2 + \frac{\alpha g \theta^2}{d\bar{T}/dz} \right] d {\bf r} \label{eq:Energy}
\end{equation}
is conserved.  It is in contrast to two-dimensional inviscid Navier Stokes equation that has two conserved quantities---the kinetic energy $\int d{\bf r} (u^2/2)$ and the enstrophy  $\int d{\bf r} ({ \nabla \!\times\! \bf u})^2/2)$~\cite{Kraichnan:PF1967}; based on these conservation laws, Kraichnan~\cite{Kraichnan:PF1967} deduced dual energy spectrum for hydrodynamic turbulence---$C_1 \Pi_u^{2/3} k^{-5/3}$ for $k < k_f$, and $C_2 \Pi_\omega^{2/3} k^{-3}$ for $k > k_f$.  Here  $\Pi_u$ and $\Pi_\omega$ are the energy flux and enstrophy flux respectively,  $k_f$ is the forcing wavenumber, and $C_1$ and $C_2$ are constants that have been estimated as 5.5--7.0~\cite{Maltrud:JFM1991,Smith:PRL1993} and 1.3--1.7~\cite{Borue:PRL1993,LINDBORG:POF2000} respectively.  Note that the stably stratified flows conserve neither the energy nor the enstrophy in the inviscid limit.

It is convenient to work with nondimensional equations  
using box height $d$ as length scale, $\sqrt{\alpha g |d\bar{T}/dz| d^2}$ as velocity scale, and $|d\bar{T}/dz|d $ as the temperature scale, which leads to
${\bf u } = {\bf u}^\prime \sqrt{\alpha g |d\bar{T}/dz| d^2}$, $\theta = \theta^\prime |d\bar{T}/dz|d$, ${\bf x} ={\bf x}^\prime d$, 
and $t = t^\prime (d/ \sqrt{\alpha g |d\bar{T}/dz| d^2})$.  Hence the nondimensionalized version of Eqs.~(\ref{eq:u_dim})-(\ref{eq:inc_dim}) are
\begin{eqnarray}
\frac{\partial {\bf u}^\prime}{\partial t^\prime} + ({\bf u}^\prime \cdot \nabla{^\prime}) {\bf u}^\prime & = & - \nabla^\prime \sigma^\prime+ \theta^\prime \hat{z} + \sqrt{\frac{\mathrm{Pr}}{\mathrm{Ra}}} \nabla{^\prime}^2 {\bf u}^\prime + {\bf f}^\prime_u, \label{eq:u_non_dim} \\
\frac{\partial \theta^\prime}{\partial t^\prime} + (\bf u^\prime \cdot \nabla^\prime) \theta^\prime & = & -  u_z^\prime +  \frac{1}{\sqrt{\mathrm{RaPr}}}\nabla{^\prime}^2 \theta^\prime, \label{eq:th_non_dim} \\
\nabla^\prime \cdot \bf u^\prime & = & 0 \label{eq:inc_non_dim}.
\end{eqnarray}
From this point onward we drop the primes on the variables for convenience.

In this paper we solve the above equations numerically, and study the energy spectra and fluxes in the regimes of strong stratification, moderate stratification, and weak stratification.   Note that the KE spectrum $[E_u(k)]$ and the potential energy (PE) spectrum $[E_{\theta}(k)]$ are defined as 
\begin{eqnarray}
E_u(k) &=& \sum_{k -1 < k^{\prime} \leq k} \frac{1}{2} |u({\bf k^\prime})|^2, \label{eq:KE_spectrum} \\
E_{\theta}(k) &=& \sum_{k -1 < k^{\prime} \leq k} \frac{1}{2} |\theta({\bf k^\prime})|^2.\label{eq:PE_spectrum} 
\end{eqnarray}

In Fourier space, the equation for the kinetic energy $E_u(k)$ of the wavenumber shell of radius $k$ is derived from Eq.~(\ref{eq:u_non_dim}) as~\cite{Verma:EPL2012,Lesieur:book}
\begin{equation}
\frac{\partial E_u(k)}{\partial t} = T_u(k) + F_B(k) + F_{ext}(k) - D(k),
\label{eq:dEk_dt}
\end{equation}
where $T_u(k)$ is the energy transfer rate to the shell $k$ due to nonlinear interaction, and $F_B(k)$ and $F_{ext}(k)$ are  the energy supply rates to the shell from the  buoyancy  and external forcing ${\bf f}_u$ respectively, i.e.,
 \begin{eqnarray}
F_B(k) & = &\sum_{|{\mathbf k}| = k} \Re(\langle u_z({\mathbf k}) \theta^*({\mathbf k}) \rangle), \\ \label{eq:FB} 
F_{ext}(k) & = & \sum_{|{\mathbf k}| = k}  \Re(\langle {\mathbf u}({\mathbf k}) \cdot {\mathbf f_u}^*({\mathbf k}) \rangle). \label{eq:Fext}
 \end{eqnarray}
In Eq.~(\ref{eq:dEk_dt}), $D(k)$ is the viscous dissipation rate  defined by
 \begin{equation}
D(k) = \sum_{|{\mathbf k}| = k} 2 \nu k^2 E_u(k).
\label{eq:D_k}
\end{equation}

The kinetic energy flux $\Pi_u(k_0)$,  which is defined as the kinetic energy leaving a wavenumber sphere of radius $k_0$ due to nonlinear interaction,  is related to the nonlinear interaction term $T_u(k)$ as 
\begin{equation}
\Pi_u(k) = - \int_0^k T_u(k)dk.
\label{eq:Pi_def}
\end{equation}
Under a steady state [$\partial E_u(k)/\partial t = 0$], using Eqs.~(\ref{eq:dEk_dt}) and (\ref{eq:Pi_def}), we deduce that
\begin{equation}
\frac{d}{dk} \Pi_u(k) =  F_B(k) + F_{ext}(k) - D(k)
\label{eq:dPik_dk}
\end{equation}
or
\begin{equation}
\Pi_u(k+\Delta k) =  \Pi_u(k) + ( F_B(k) + F_{ext}(k) - D(k)) \Delta k.
\label{eq:Pi_k_Fk_Dk}
\end{equation}

In computer simulations, the KE flux, $\Pi_u(k_0)$,  is computed by the following formula~\cite{Dar:PD2001,Verma:PR2004},
\begin{equation}
\Pi_u(k_0)  =  \sum_{k > k_0} \sum_{p\leq k_0} \delta_{\bf k,\bf p+ \bf q} \Im([{\bf k \cdot u(q)}]  [{\bf u^*(k) \cdot u(p)}]). \label{eq:ke_flux}
\end{equation}
Similarly, the enstrophy flux $[\Pi_{\omega}(k_0)]$ and PE flux $[\Pi_{\theta}(k_0)]$ are the enstrophy and the potential energy leaving a wavenumber sphere of radius $k_0$ respectively. The formulae to compute these quantities are
\begin{eqnarray}
\Pi_{\omega}(k_0)  & = & \sum_{k > k_0} \sum_{p \leq k_0} \delta_{\bf k,\bf p+ \bf q} \Im([{\bf k \cdot u(q)}]  [{ {\omega}^*(\bf k)  {  \omega}(\bf p)}]), \label{eq:ens_flux} \\
\Pi_{\theta}(k_0)  & = &  \sum_{k > k_0} \sum_{p \leq k_0} \delta_{\bf k,\bf p+ \bf q} \Im([{\bf k \cdot u(q)}]  [{ {\theta}^*(\bf k) \theta(p)}]). \label{eq:pe_flux} 
\end{eqnarray}
Note that the total energy flux $\Pi_{\mathrm{Total}}(k)$ is defined as 
\begin{equation}
\Pi_{\mathrm{Total}}(k) = \Pi_u(k) + \Pi_{\theta}(k).\label{eq:total_flux}
\end{equation}
In the above expression, the prefactors are unity due to nondimensionalization.

We will compute the aforementioned spectra and fluxes  using the steady-state numerical data.

\section{Simulation method}
\label{sec:sim_method}
We solve Eqs.~(\ref{eq:u_non_dim})-(\ref{eq:inc_non_dim}) numerically using a pseudo-spectral code Tarang~\cite{Verma:Pramana2013}.    We employ the fourth-order Runge-Kutta method  for time stepping,  the Courant-Friedrichs-Lewy (CFL) condition to determines the time step $\Delta t$, and $2/3$ rule for dealiasing.   We use periodic boundary conditions  on both sides of a square box of dimension  $2\pi \times 2\pi$. Since the system is stable, we apply random large-scale forcing in the band $2 \leq k \leq 4$ to obtain a stastically-steady turbulent flow. For details of numerical method, refer to~\cite{Verma:Pramana2013,Kumar:PRE2014}.

\setlength{\tabcolsep}{5pt}
\begin{table}[htbp]
\begin{center}
\caption{Parameters of our direct numerical simulations (DNS): Froude number $\mathrm{Fr}$; grid resolution; Rayleigh number $\mathrm{Ra}$; energy supply rate $\varepsilon$; Richardson number $\mathrm{Ri}$; Reynolds number $\mathrm{Re}$; kinetic energy dissipation rate $\epsilon_u$; potential energy dissipation rate $\epsilon_{\theta}$; anisotropy ratio $A = {\langle u^2_\perp \rangle}/{\langle u^2_\parallel \rangle}$; and $k_{max}\eta$, where $k_{max}$ is the maximum wavenumber and $\eta$ is the Kolmogorov length. We kept the Prandtl number $\mathrm{Pr} = 1$ for all our runs.}
\hspace{40mm}
\begin{tabular}{c  c c  c  c c   c  c  c  c}
\hline \hline \\
$\mathrm{Fr}$ & Grid &$\mathrm{Ra}$ & $\varepsilon$ & $\mathrm{Ri}$ & $\mathrm{Re}$   &  $\epsilon_u$ & $\epsilon_{\theta}$ & $A$ & $k_{max} \eta$ \\[2 mm]
\hline \\
$0.16$  &$512^2$ & $10^9$ & $10^{-6}$ & $40$ &$5 \times 10^3$  &$2.2 \times 10^{-6}$ & $1.7 \times 10^{-7}$ & $38$ &$2.8$\\

$0.31$  & $512^2$ & $10^8$ & $10^{-4}$ & $11$ &$3.1 \times 10^3$  & $2.2 \times 10^{-5}$ & $6.2 \times 10^{-6}$ & $38$ &$3.7$  \\

$0.37$  & $8192^2$ & $10^{10}$ & $0.01$ & $7.3$ &$3.7 \times 10^4$  & $4.9 \times 10 ^{-4}$ & $2.8 \times 10^{-3}$ & $3.4$ & $4.9$ \\

$0.45$  & $512^2$ & $10^8$ & $0.01$ & $4.9$ &$4.5 \times 10^3$  & $1.2 \times 10^{-3}$ & $2.9 \times 10^{-3}$ & $4.2$ &$1.4$  \\

$0.73$  & $2048^2$ & $10^8$ & $0.1$ & $1.9$ &$7.3 \times 10^3$  & $5.3 \times 10^{-3}$ & $2.7 \times 10^{-2}$ & $1.6$ &$3.8$ \\

$1.1$  & $2048^2$ & $10^8$ & $0.3$ & $0.8$ &$1.1 \times 10^4$  & $9.2 \times 10^{-3}$ & $7.7 \times 10^{-2}$ & $1.4$ &$3.3$  \\
\hline 

\end{tabular}
\label{table:simulation_details}
\end{center}
\end{table}

\begin{figure}[htbp]
\begin{center}
\includegraphics[scale = 1]{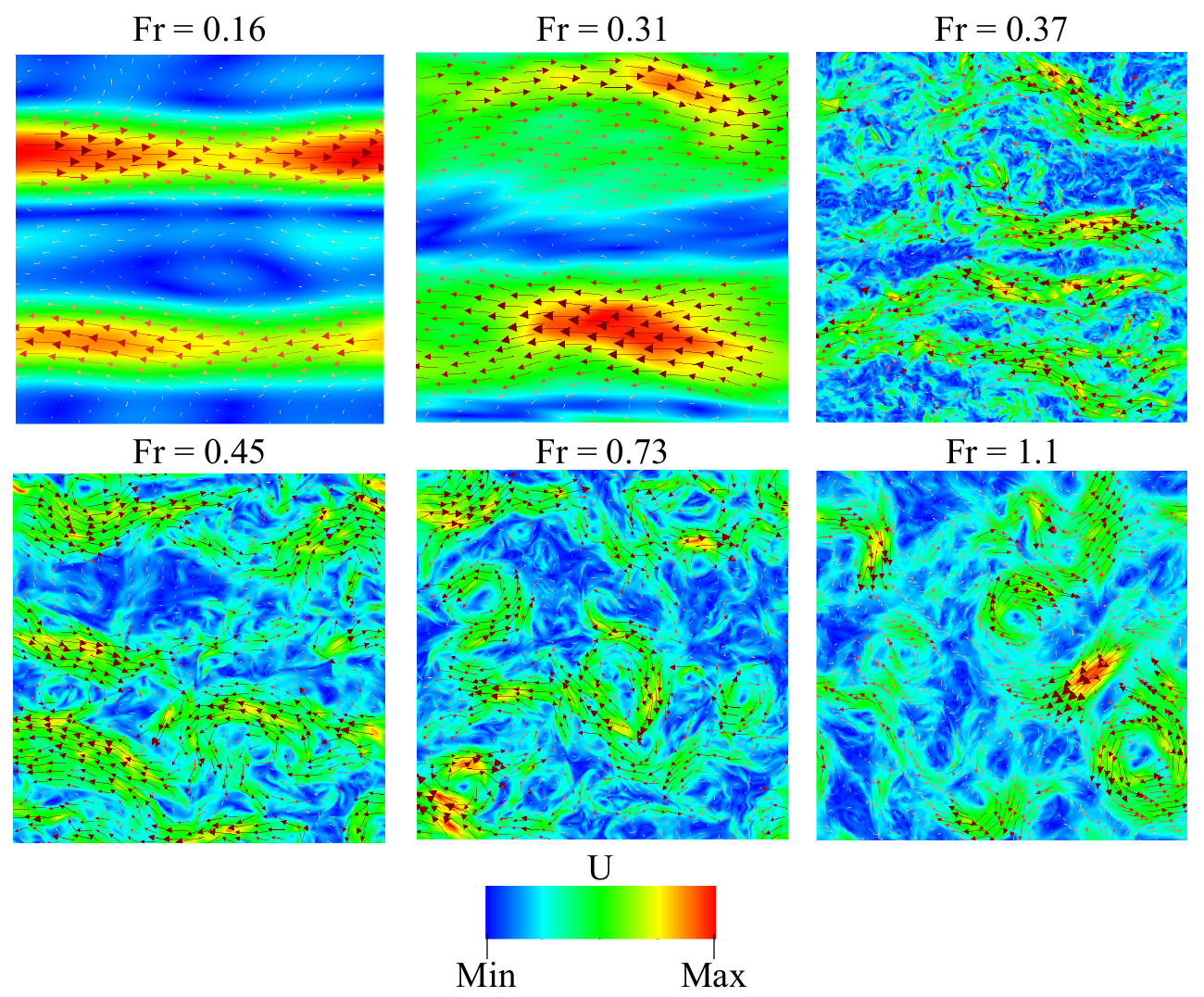}
\end{center}
\setlength{\abovecaptionskip}{0pt}
\caption{For $\mathrm{Fr} = 0.16$, $0.31$, $0.37$, $0.45$, $0.73$, and $1.1$, the density plots of the magnitude of velocity field with the velocity vector $\bf u$  superposed on it. For low $\mathrm{Fr}$,  fluctuations are suppressed along buoyancy direction.  However they grow gradually on the increase of $\mathrm{Fr}$.  We classify $\mathrm{Fr}=0.16,0.31$ as strong stratification, 0.37  and 0.45 as moderate stratification,  1.1 as weak stratification, and 0.73 as transition between moderate and weak stratification.}
\label{fig:density_all}
\end{figure}

In Table~\ref{table:simulation_details} we list the set of parameters for which we performed our simulations.  We employ grid resolutions of $512^2$ to $8192^2$, the higher ones for higher Reynolds number.  The Rayleigh number of our simulations ranges from $10^8$ to $10^{10}$, while the Reynolds number ranges from $5000$ to $3.7\times 10^4$.  All our simulations are fully resolved since $k_{max}\eta > 1$, where $\eta$ is the  Kolmogorov length scale, and $k_{max}$ is the maximum wavenumber attained in DNS for a particular grid resolution.  Note that the energy supply rate, $\varepsilon$, is greater than the viscous dissipation rate, $\epsilon_u$, with the balance getting transferred to the potential energy via buoyancy ($\alpha g \theta u_z$).

The Froude number of our simulations are $\mathrm{Fr} = 0.16$, $0.31$, $0.37$, $0.45$, $0.73$, and $1.1$; the lowest $\mathrm{Fr}$ correspond to the strongest stratification, while the largest $\mathrm{Fr}$ to the weakest stratification.  We show in subsequent discussion that the flow behaviour in these regimes are very different.  One of the major difference is the anisotropy that is quantified using an anisotropy parameter $A = {\langle u^2_\perp \rangle}/{\langle u^2_\parallel \rangle}$. For the strongest stratification with $\mathrm{Fr} = 0.16$, $A \approx 38$ indicating a strong anisotropy.  However, for the weakest stratification with $\mathrm{Fr} = 1.1$, $A \approx 1.4$, indicating a near isotropy.  

\section{Results}
We begin with a qualitative description of the flow profiles for the strongly, moderately, and weakly stratified regimes. 
In Fig.~\ref{fig:density_all} we show the velocity vectors superposed on the magnitude of the velocity field. 
For strong stratification ($\mathrm{Fr} = 0.16$), 
we observe two robust flow structures moving in the opposite directions, i.e., a VSHF~\cite{SW:JFM}. 
On further increasing $\mathrm{Fr}$ to 0.31, the streams widen and start to diffuse, and the  flow becomes more random. 
In the moderately stratified regime ($\mathrm{Fr}=0.37,0.45$), the streams break into filaments, and the flow becomes progressively
disordered.
Finally, for weak stratification ($\mathrm{Fr}=1.1$), the flow appears turbulent during which the aforementioned filaments 
tend to be wrapped into compact isolated vortices~\cite{McWilliams}.  The transition from moderate to weak stratification occur near $\mathrm{Fr} = 0.73$.

\subsection{Strong Stratification}
\label{sec:gravity_wave}

Here we focus on the simulation with $\mathrm{Fr}=0.16$, $\mathrm{Re}=5000$, $\mathrm{Pr}=1$, $\mathrm{Ra}=10^9$, and forcing
amplitude $\varepsilon = 10^{-6}$.
The flow exhibits strong anisotropy as is evident from the ratio $A = {\langle u^2_\perp \rangle}/{\langle u^2_\parallel \rangle} = 38$.
In fact, the flow exhibits wave-like behaviour that can be confirmed by studying the dominant
Fourier modes.

\begin{figure}[htbp]
\begin{center}
\includegraphics[scale = 0.75]{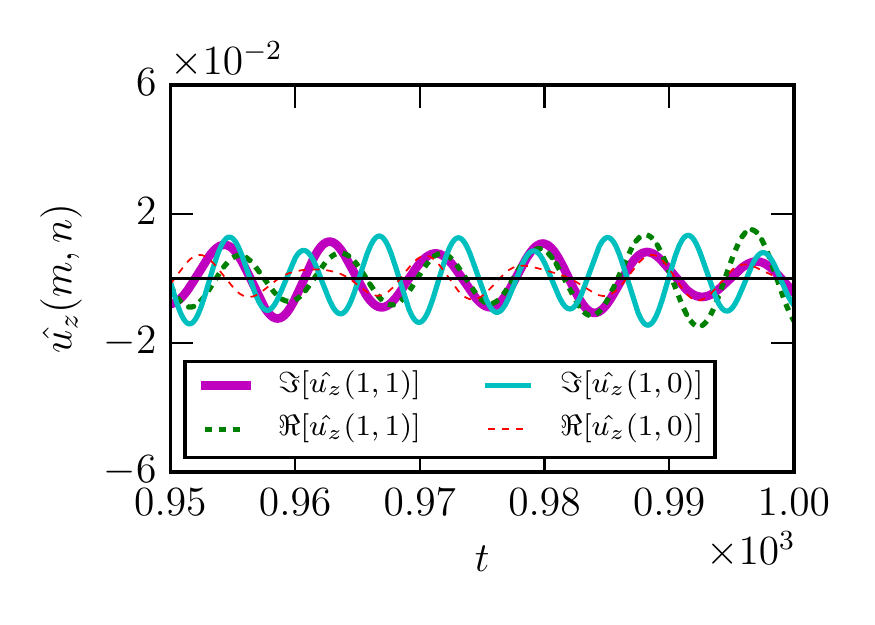}
\end{center}
\setlength{\abovecaptionskip}{0pt}
\caption{Time series of the real and imaginary parts of Fourier modes $\hat{u}_z(1,1)$ and $\hat{u}_z(1,0)$ for strong stratification ($\mathrm{Fr}=0.16$).}
\label{fig:mode}
\end{figure}

We compute the most energetic Fourier modes in the flow and find the modes $(1,0)$ and $(1,1)$ to be the most dominant. Fig.~\ref{fig:mode} shows the time series of
the real and imaginary parts of $\hat{u}_z(1,1)$ and $\hat{u}_z(1,0)$ from which we extract the oscillation time period of these modes as approximately $8.4$ and $6.5$,
and their frequencies as $0.75$ and $0.97$.
These numbers match very well with the dispersion relation [Eq.~(\ref{eq:dispersion})], thus we interpret these structures to be internal
gravity waves.
Note that these robust flow structures moving in
horizontal directions constitute the VSHF.
We also observe that  $\hat{\bf u}(0,n) \approx 0$ where $n$ is an integer, so almost no energy is transferred to purely zonal flows.
A natural vertical length-scale that emerges in strongly stratified flows is $U/f$, where $U$ is the magnitude of the horizontal
flow and $f=N/2\pi$~\cite{Billant}. With the present parameters, we see that this leads to VSHFs of size $\sim 1$, which is in reasonable
agreement with the bands seen in the first panel of Fig.~\ref{fig:density_all}.

\begin{figure}[htbp]
\begin{center}
\includegraphics[scale = 0.75]{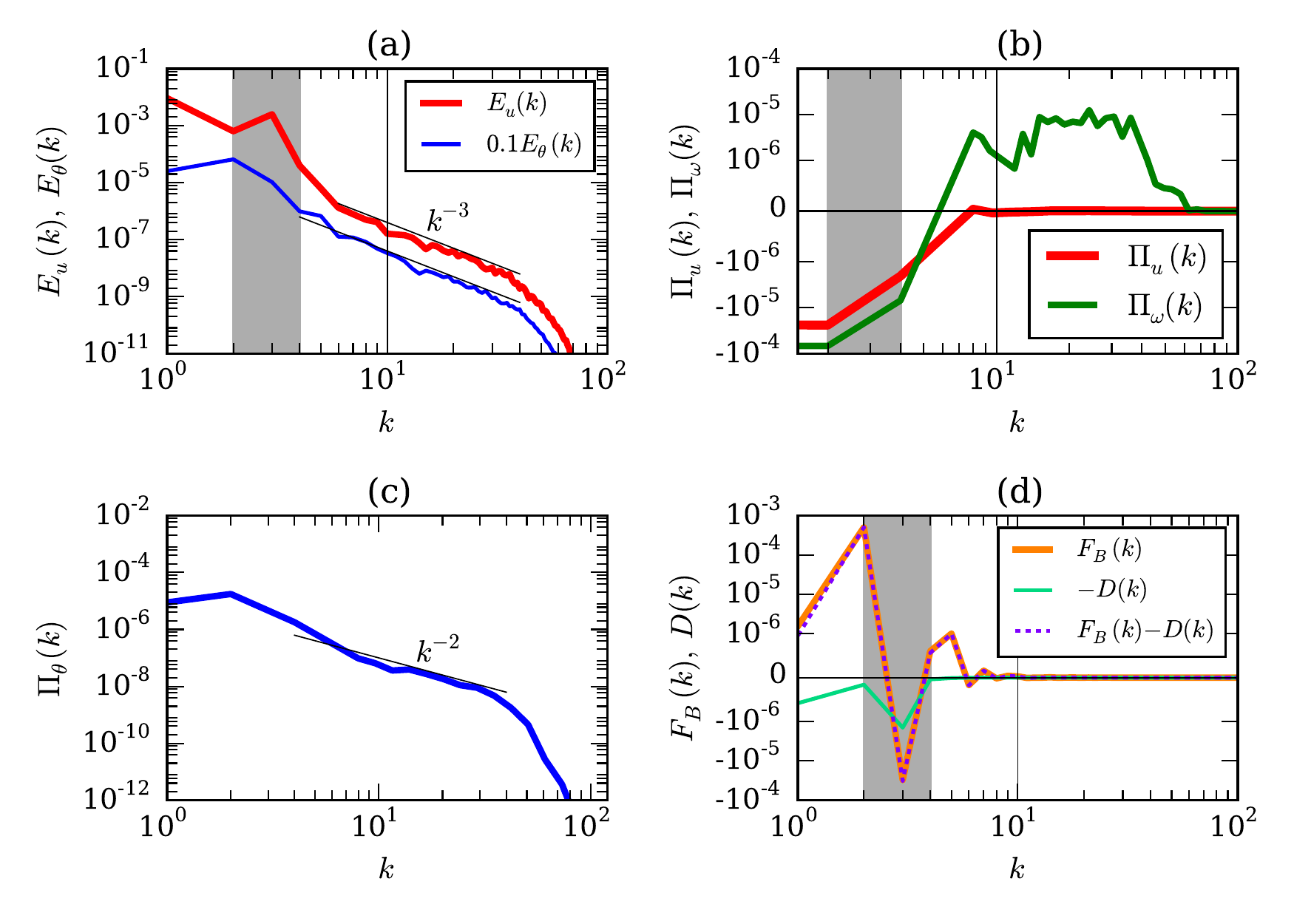}
\end{center}
\setlength{\abovecaptionskip}{0pt}
\caption{Strong stratification ($\mathrm{Fr}=0.16$): (a) KE and PE spectra;
(b) KE flux $\Pi_u(k)$ and enstrophy flux $\Pi_{\omega}(k)$. The grey shaded region shows the forcing band. The KE flux is zero for wavenumber $k \geq 10$; (c) PE flux $\Pi_{\theta}(k)$; (d) $F_B(k)$, $D(k)$, and $F_B(k)-D(k)$.}
\label{fig:low_Fr_flux}
\end{figure}

To explore the flow properties further, we compute the KE spectrum $E_u(k)$, and the KE and enstrophy fluxes. 
As shown in Fig.~\ref{fig:low_Fr_flux}(a), the KE at small scales (large $k$) is several orders of magnitude lower than that for low-$k$ modes. 
Thus, even though small-scale turbulence is present in the system, the energy content of the large scale internal gravity waves is much 
larger than the sea of small-scale turbulence. The flux computations show that the KE flux $\Pi_u(k) \approx 0$ for $k>k_f$, but the 
enstrophy flux $\Pi_{\omega}(k)$ is positive and fairly constant [see  Fig.~\ref{fig:low_Fr_flux}(b)]. For these band of wavenumbers we observe that 
\begin{equation}
E_u(k) \approx 1.0 \Pi_\omega^{2/3} k^{-3},
\label{eq:Ek_strong_stratification}
\end{equation}
which is similar to the forward enstrophy cascade regime of 2D turbulence (including the prefactor)~\cite{Borue:PRL1993,LINDBORG:POF2000}. 
The aforementioned flux computations are also consistent with the fluxes reported for 2D 
turbulence~\cite{Boffetta:JFM2007,Boffetta:PRE2010,Boffetta:ARFM2012}.

In addition, we observe that the PE spectrum, $E_\theta(k)$, scales approximately as $k^{-3}$ and the PE flux follows $\Pi_\theta(k) \sim k^{-2}$
[see Fig.~\ref{fig:low_Fr_flux}(a,c)]. The $k^{-3}$  scaling of the PE is in sharp contrast to the $k^{-1}$ Batchelor spectrum
for a passive scalar in the 2D hydrodynamic turbulence in the wavenumber regimes with a forward enstrophy cascade~\cite{Jul,Batchelor:JFM1959a}.
Further, $\Pi_\theta(k)$ decreases rapidly with wavenumber, rather than being a constant as for a passive scalar.
We demonstrate the consistency among these scalings of KE and PE as follows:
the KE spectrum $E_u(k) \sim k^{-3}$ implies $u_k \sim k^{-1}$, substitution of which in the PE flux equation yields
 \begin{equation}
\Pi_\theta \approx k u_k \theta_k^2 \sim k^{-2}.
\end{equation}
Consequently, $\theta_k \sim k^{-1}$, and hence
 \begin{equation}
E_{\theta}(k) \approx \frac{ \theta_k^2}{k} \sim  k^{-3}.
\end{equation}
Also,  Fig.~\ref{fig:low_Fr_flux}(d) shows the energy supply rate due to buoyancy $F_B(k)$ and the dissipation rate $D(k)$. 
The buoyancy is active at large-scales only, and it is quite small for $k \geq10$. 

\begin{figure}[htbp]
\begin{center}
\includegraphics[scale = 0.8]{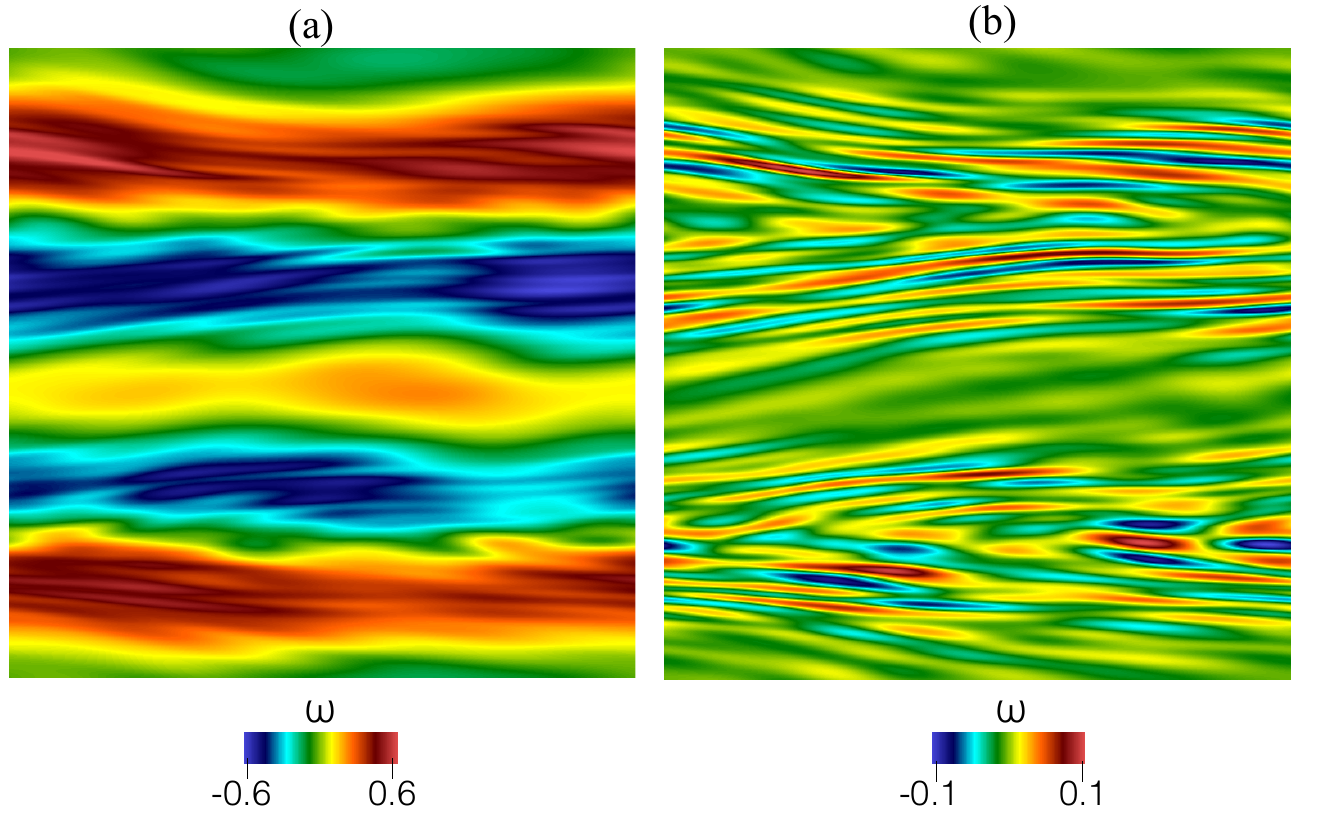}
\end{center}
\setlength{\abovecaptionskip}{0pt}
\caption{For $\mathrm{Fr}=0.16$, (a) density plot of the magnitude of vorticity field $\omega$; (b)  The density plot of  the  vorticity field $\omega$; for this we  truncate the modes in the wavenumber band $0\leq k \leq 10$.}
\label{fig:low_Fr_vorticity}
\end{figure}

So, for strong stratification, the picture that emerges is of large-scale internal gravity waves, physically manifested as a VSHF,
that co-exist with small scale turbulence which is characterized by an approximate $k^{-3}$ scaling for both the KE and PE. Moreover, the KE flux is close
to zero while the PE flux is positive and closely follows a $k^{-2}$ power-law. Thus, the total energy of the system is
systematically transferred to small scales. 
To visualize the coexistence of the large-scale internal gravity waves and small scale turbulence we plot the vorticity field
in Fig.~\ref{fig:low_Fr_vorticity}(a), and the small-scale vorticity field after removing the large-scale wavenumbers from the
band $0\leq k \leq 10$ in Fig.~\ref{fig:low_Fr_vorticity}(b). Clearly we observe large-scale internal gravity waves or VSHF riding on a
sea of small-scale turbulence.

\subsection{Moderate Stratification}
\label{sec:bolgiano}
Among our simulations, the density stratification is moderate for $\mathrm{Fr}=0.37$ and 0.45 (see Fig.~\ref{fig:density_all}).
In this subsection we will focus on  $\mathrm{Fr}=0.37$ which is obtained
for $\mathrm{Ra}=10^{10}$ and an energy supply rate of $\varepsilon = 0.01$. For this case,  $\mathrm{Re} = 3.7 \times 10^4$.
As shown in Fig.~\ref{fig:density_all}, the flow pattern for the above set of parameters differs significantly from that corresponding to strong
stratification. In fact, there is no evidence of a VSHF in Fig.~\ref{fig:density_all} for $\mathrm{Fr}=0.37$.

With regard to turbulence phenomenology, Bolgiano~\cite{Bolgiano:JGR1959} and Obukhov~\cite{Obukhov:DANS1959} (denoted by BO) were among the first to 
consider stably stratified flows in 3D. According to this phenomenology, for large scales, i.e., $k< k_B$,
\begin{eqnarray}
E_u(k) & =  & c_1 (\alpha^2 g ^2 \epsilon_\theta)^{2/5}k^{-11/5}, \label{eq:Eu} \\
E_\theta(k) & =  & c_2 (\alpha g)^{-2/5}\epsilon_\theta^{4/5} k^{-7/5}, \label{eq:Etheta} \\
\Pi_u(k) & = & c_3 (\alpha^2 g^2 \epsilon_\theta)^{3/5} k^{-4/5},  \label{eq:pi} \\
\Pi_\theta(k) & = &  \epsilon_\theta = \mathrm{constant}, \label{eq:pi_theta}  \\
k_B & = &  c_4 (\alpha g)^{3/2}\epsilon_u^{-5/4}\epsilon_{\theta}^{3/4}, \label{eq:k_B}  
\end{eqnarray}
where $c_i$'s are constants, $\epsilon_u$ is the kinetic energy supply rate, $\epsilon_\theta$ is the potential energy supply rate, 
and $k_B$ is the Bolgiano wavenumber. At smaller scales ($k>k_B$), BO argued that the buoyancy effects are weak, and hence Kolmogorov's spectrum is valid in this regime, i.e.,
\begin{eqnarray}
E_u(k) & =  & C_u  \epsilon_u^{2/3}k^{-5/3}, \label{eq:Eu_KO} \\
E_\theta(k) & =  & C_{\theta} \epsilon_u^{-1/3}\epsilon_\theta k^{-5/3}, \label{eq:Etheta_KO} \\
\Pi_u(k) & = &  \epsilon_u = \mathrm{constant},  \label{eq:pi_KO} \\
\Pi_\theta(k) & = &  \epsilon_\theta = \mathrm{constant}, \label{eq:pi_theta_KO}
\end{eqnarray}
where $C_u$ and $C_{\theta}$ are Kolmogorov's and Batchelor's constants respectively. 
Note that the recent developments~\cite{Kumar:PRE2014,Kumar:PRE2015,Verma:PF2015,Bhattacharjee:PLA2015} in 3D SS turbulence have confirmed 
the existence of Bolgiano scaling. 

For 2D SS turbulence, Eqs.~(\ref{eq:Eu_KO}-\ref{eq:pi_theta_KO}) need to be modified for the $k>k_B$ regime since 2D hydrodynamic turbulence yields $k^{-3}$ energy spectrum at small scales due to constant enstrophy cascade.  Thus, modifications of Eqs.~(\ref{eq:Eu_KO}-\ref{eq:pi_theta_KO}) take the form,
\begin{eqnarray}
E_u(k) & =  & c_5  \Pi_{\omega}^{2/3}k^{-3}, \label{eq:Eu_KO_2D} \\
E_\theta(k) & =  & c_6 k^{-1}, \label{eq:Etheta_KO_2D} \\
\Pi_\omega(k) & = &  \epsilon_\omega = \mathrm{constant},  \label{eq:pi_KO_2D} \\
\Pi_\theta(k) & = &  \epsilon_\theta = \mathrm{constant}. \label{eq:pi_theta_KO_2D}
\end{eqnarray}
Here $\Pi_\theta(k) \sim k u_k \theta_k^2 = \mathrm{const}$. As $u_k \sim k^{-1}$, this yields $\theta_k \sim \mathrm{const}$.
Hence, we argue that $E_\theta(k) \sim \theta_k^2/k \sim k^{-1}$. At these smaller scales, it is important to note that the degree
of nonlinearity is expected to
be higher for moderate stratification, which leads to $E_\theta(k) \sim k^{-1}$ and $\Pi_\theta(k) \sim \mathrm{const}$, in contrast to
$E_\theta(k) \sim k^{-3}$ and $\Pi_\theta(k) \sim k^{-2}$ for strongly stratified flows.

\begin{figure}[htbp]
\begin{center}
\includegraphics[scale = 0.75]{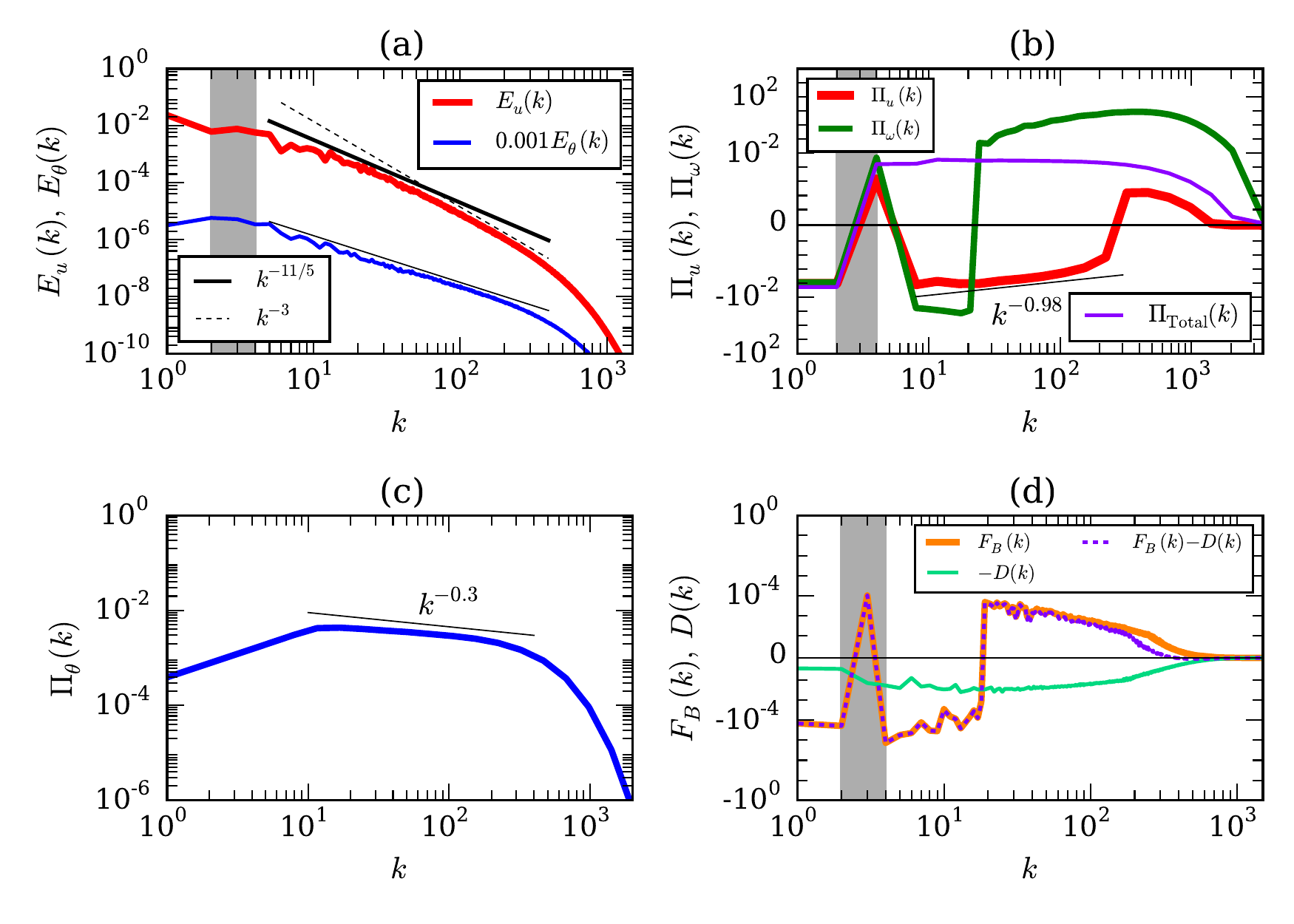}
\end{center}
\setlength{\abovecaptionskip}{0pt}
\caption{Moderate stratification ($\mathrm{Fr}=0.37$): (a) The KE and PE spectra. KE spectrum shows dual scaling with $k^{-11/5}$ and $k^{-3}$. The best fit  for PE spectrum is $k^{-1.64}$ (thin black line); (b) KE flux $\Pi_u(k)$, enstrophy flux $\Pi_{\omega}(k)$, and total energy flux $\Pi_{\mathrm{Total}}(k)$; (c) PE flux; (d) $F_B(k)$, $D(k)$, and $F_B(k)-D(k)$.}
\label{fig:mid_Fr_spectrum}
\end{figure}

The KE and PE spectra as well as their fluxes are shown in Fig.~\ref{fig:mid_Fr_spectrum}. The KE spectrum $E_u(k)$ exhibits BO scaling,
in particular, $E_u(k) \sim k^{-11/5}$ for $5 \leq k \leq 90$, and $E_u(k) \sim k^{-3}$ for  $90 \leq k \leq 400$.
The KE flux, seen in Fig.~\ref{fig:mid_Fr_spectrum}(b), also varies with scale; at large scales we
see an inverse transfer (that scales approximately as $k^{-0.98}$), while at small scales we obtain a forward transfer of KE. The enstrophy flux is
positive except for a narrow band near $k \approx 10$.
Note that the PE spectrum [Fig.~\ref{fig:mid_Fr_spectrum}(a)] does not show dual scaling and scales approximately as
$k^{-1.64}$, its flux is also not a constant but follows $\Pi_\theta(k) \sim k^{-0.3}$. 

At large scales, these scaling laws can be explanied by replacing $\epsilon_\theta$ of
Eqs.~(\ref{eq:Eu}-\ref{eq:pi}) with $\Pi_{\theta}(k) \sim k^{-0.3}$. This is similar to the
variable flux arguments presented by Verma~\cite{Verma:EPL2012} and Verma and Reddy~\cite{verma:POF2015}.
Specifically,
\begin{eqnarray}
E_u(k) & \sim & k^{-0.3 \times 2/5}k^{-11/5} \sim k^{-2.32}, \label{eq:Eu_new} \\
E_{\theta}(k) & \sim &  k^{-0.3 \times 4/5}k^{-7/5} \sim k^{-1.64}, \label{eq:Etheta_new} \\
\Pi_u(k) & \sim & k^{-0.3 \times 3/5}k^{-4/5} \sim k^{-0.98}. \label{eq:Piu_new}
\end{eqnarray}
Indeed, the spectral indices obtained above match those in Fig.~\ref{fig:mid_Fr_spectrum} very closely.

At small scales the KE spectrum $E_u(k) \approx 2.0 \Pi_\omega^{2/3} k^{-3}$ that is associated with weaker buoyancy and constant
enstrophy flux [see Fig.~\ref{fig:mid_Fr_spectrum}(a,b)], and in is accord with  Eq.~(\ref{eq:Eu_KO_2D}).
As mentioned, the PE spectrum does not exhibit dual scaling, and we do not see the $k^{-1}$ scaling expected from Eq.~(\ref{eq:Etheta_KO_2D})
at small scales. Though, it should be noted that the PE flux is not constant but scales approximately as $k^{-0.3}$, and this implies
a slighltly steeper ($k^{-4/3}$) small scale PE spectrum. Indeed, a small change in scaling of this kind, i.e., $-1.64$ and $-1.33$ at large and small scales, respectively,
may only be observable at a higher resolution.

\begin{figure}[htbp]
\begin{center}
\includegraphics[scale = 0.9]{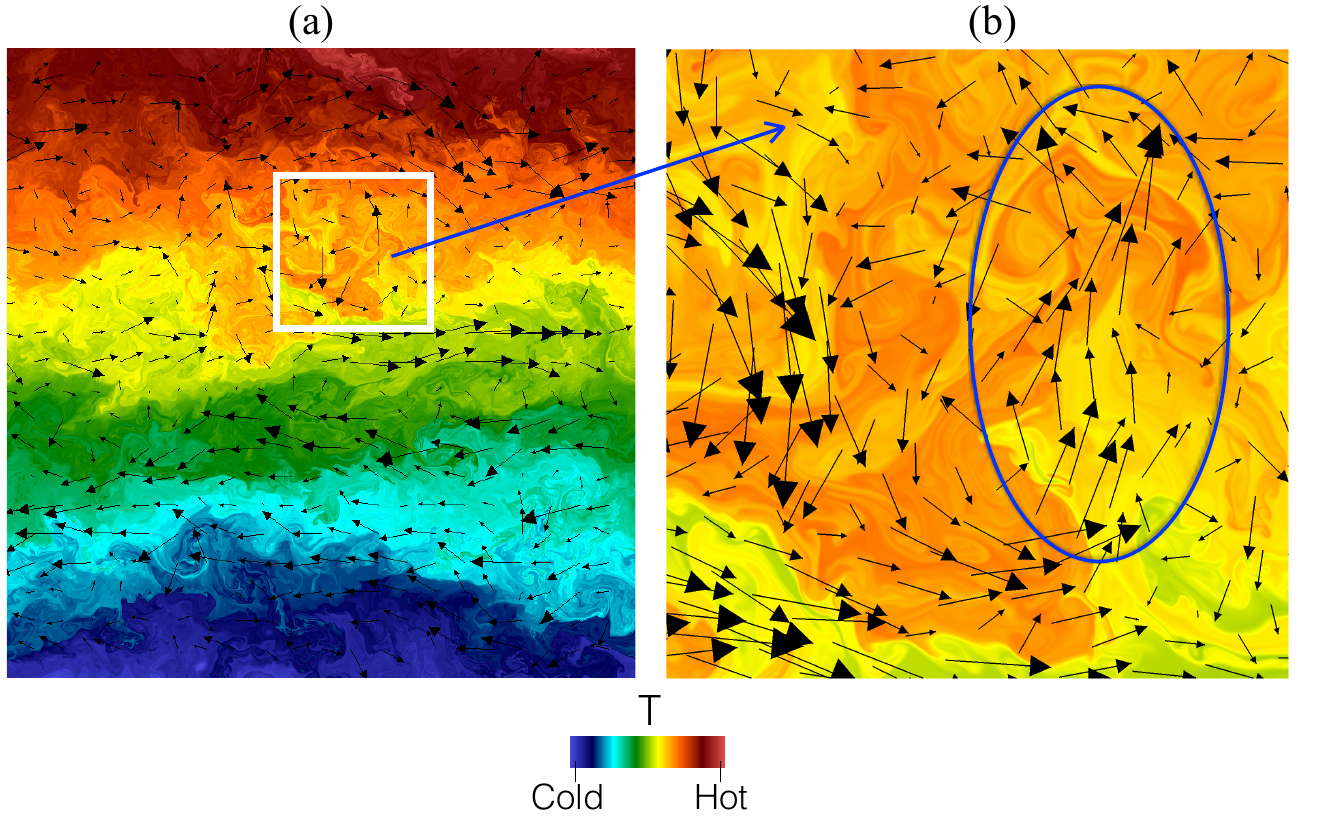}
\end{center}
\setlength{\abovecaptionskip}{0pt}
\caption{For $\mathrm{Fr}=0.37$ (a) the density plot of the temperature field with the velocity field  superimposed on it. In the boxed zone, hotter (higher) fluid ascends 
 thus $F_B(k) \propto u_z \theta >0 $; (b) a zoomed view of the boxed zone.}
\label{fig:mid_Fr_density}
\end{figure}

The KE flux in the present 2D setting exhibits an inverse cascade, in contrast to the forward
cascade in 3D~\cite{lind}. Still the $k^{-11/5}$ spectrum of BO scaling is valid in 2D SS turbulence due to the following
reason. The energy supply due to buoyancy $F_B(k)$ and the dissipation rate $D(k)$,
shown in Fig.~\ref{fig:mid_Fr_spectrum}(d), exhibit $F_B(k)>0$ for $k > 20$, in contrast to 3D SS flows for which  $F_B(k) < 0$~\cite{Kumar:PRE2014}.
From  Eq.~(\ref{eq:Pi_k_Fk_Dk}) we deduce that $|\Pi_u(k+\Delta k)| < |\Pi_u(k)|$ when $\Pi_u(k) < 0$ and $F_B(k)>0$. Thus $|\Pi_u(k)|$
decreases with $k$ and this yields Bolgiano scaling for the 2D moderately stratified flows.
Physically, in Fig.~\ref{fig:mid_Fr_density} we observe ascending hot fluid for which $u_z$ and $\theta$ are positively correlated.
This is in contrast to 3D SS flows for which $F_B(k) < 0$ due to a conversion of KE to PE~\cite{Kumar:PRE2014}; i.e., there $u_z$ and $\theta$
are anti-correlated. Finally, it should be noted that even though KE flows upscale in this 2D setting, the total energy is transferred from
large to small scales as seen by the total energy flux $\Pi_{\mathrm{Total}}(k)$ in Fig.~\ref{fig:mid_Fr_spectrum}(b).

\subsection{Weak Stratification}
\label{sec:weak_stratification}

Lastly we discuss the flow behaviour for weak stratification. In our simulations this is achieved for  $\mathrm{Ra}=10^8$, $\varepsilon=0.3$  that yields $\mathrm{Fr}=1.1$ and $\mathrm{Re}  = 1.1 \times 10^4$. The flow pattern in Fig.~\ref{fig:density_all} for $\mathrm{Fr}=1.1$ shows a  complete lack of a VSHF, instead, there is  a tendency to form isotropic coherent structures similar to 2D hydrodynamic turbulence~\cite{McWilliams}. 

\begin{figure}[htbp]
\begin{center}
\includegraphics[scale = 0.75]{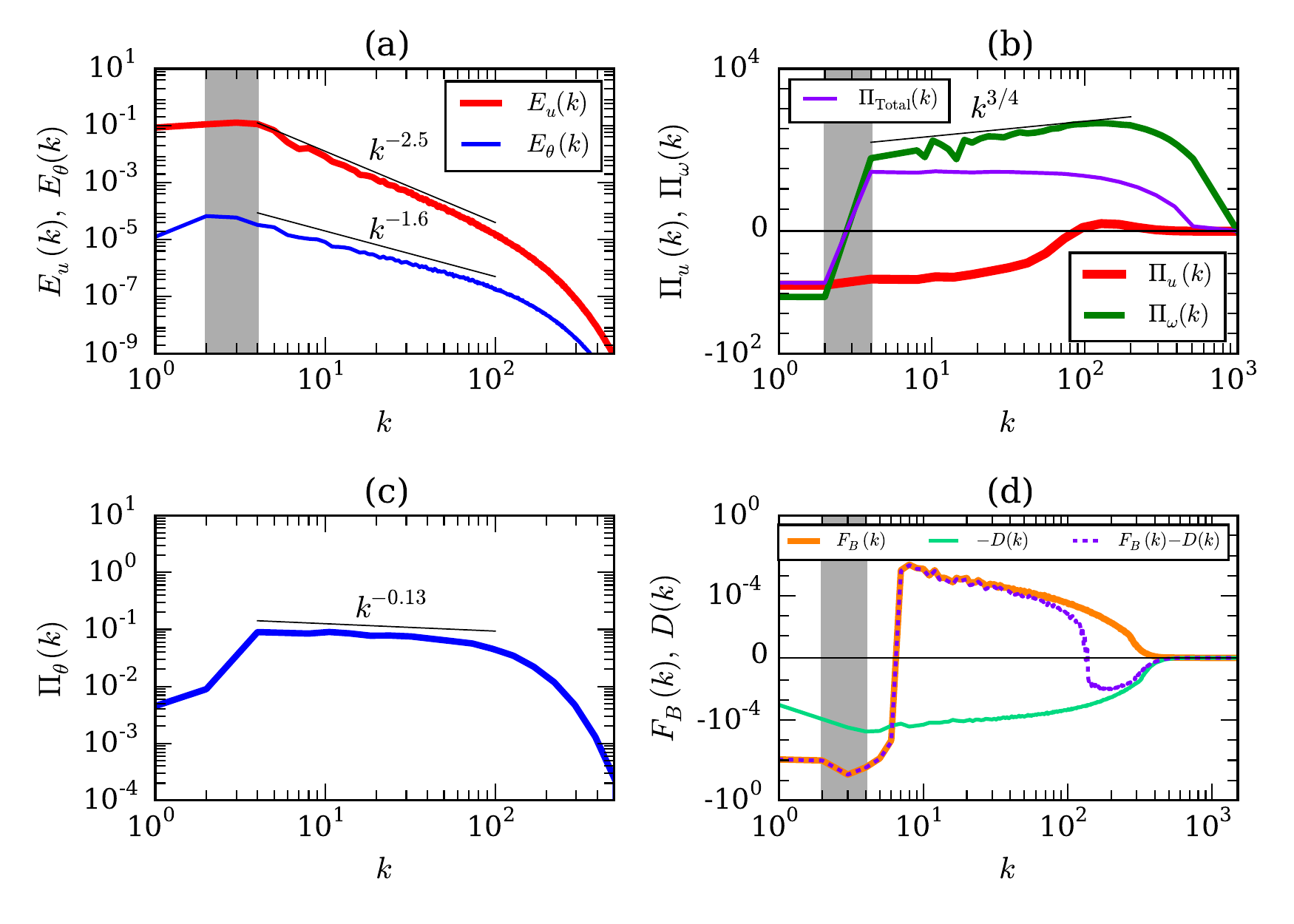}
\end{center}
\setlength{\abovecaptionskip}{0pt}
\caption{Weak stratification ($\mathrm{Fr}=1.1$): (a) Plots of KE and PE spectra. KE spectrum $E_u(k)$ shows $k^{-2.5}$ scaling, while PE spectrum $E_{\theta}(k)$ shows $k^{-1.6}$ scaling; (b) Plots of KE flux $\Pi_u(k)$, enstrophy flux $\Pi_{\omega}(k)$, and total energy flux $\Pi_{\mathrm{Total}}(k)$; (c) PE flux $\Pi_{\theta}(k)$; (d) $F_B(k)$, $D(k)$, and $F_B(k)-D(k)$.}
\label{fig:High_Fr_spectrum}
\end{figure}

The energy spectra and fluxes for this case are shown in Fig.~\ref{fig:High_Fr_spectrum}.
Qualitatively similar to the moderate stratification case, we observe a negative KE flux at large scales.
The enstrophy flux is strong and always positive, in fact it increases with wavenumber
and scales approximately
as $k^{3/4}$ up to the dissipation scale. This feature
of the enstrophy flux alters the energy spectrum as follows:

\begin{equation}
E_u(k) \approx {\Pi_\omega}^{2/3} k^3 \sim k^{3/4 \times 2/3 -3} \sim k^{-2.5},
\label{aa}
\end{equation}
which is in good agreement with our numerical finding, as shown in Fig.~\ref{fig:High_Fr_spectrum}(a).

The PE spectrum $E_\theta(k) \sim k^{-1.6}$ and its flux $\Pi_\theta$ is approximately constant with $\Pi_\theta(k) \sim k^{-0.13}$.
Using $E_u(k) \sim k^{-2.5}$ and $\Pi_\theta(k) \approx ku_k \theta_k^2 \sim k^{-0.13}$ , we obtain
$E_\theta(k) \approx {\theta_k}^2/k \sim k^{-1.38}$, which is a little shallower than the $k^{-1.6}$ spectrum obtained in our
numerical simulation.
Once again, the total energy in the system
flows from large to small scales [see Fig.~\ref{fig:High_Fr_spectrum}(b)]. Taken together, the behaviour of KE and PE suggest that BO scaling may still be applicable for $\mathrm{Fr}=1.1$,
though with a very restricted shallow ($-11/5$) large-scale KE spectrum. Indeed, it would require much higher resolution to probe this
issue.
Finally, we remark that, to some extent, the forward (inverse) transfer of PE (KE) is reminiscent
the flux loop scenario proposed by Boffetta {\em et al.}~\cite{Boffetta:EPL2011} for weakly stratified flows.

\section{Summary and conclusions}
\label{sec:conclusion}

We performed direct numerical simulations of 2D stably stratified flows under large-scale random forcing and studied the spectra and 
fluxes of kinetic energy, enstrophy, 
and potential energy. The flows exhibit different behaviour as the strength of stratification is varied, and this is 
summarized in Table~\ref{table:summary}. 

\setlength{\tabcolsep}{0.05pt}
\begin{table}[htbp]
\begin{center}
\caption{Scaling of KE spectrum $E_u(k)$, PE spectrum $E_{\theta}(k)$, KE flux $\Pi_u(k)$, PE flux $\Pi_{\theta}(k)$, and enstrophy flux $\Pi_{\omega}(k)$ for diffrent strength of stratification.}
\hspace{40mm}

\begin{tabular}{c | c | c}
    \hline
    Strength of stratification & Spectrum &Flux \\ \hline
    \multirow{6}{*}{Strong}&Large scale VSHF & \\ &   & \\ & & \\  & Small scale turbulence:  & \\ &$E_u(k) \sim k^{-3}$ & $\Pi_{u}(k)\sim 0$\\ &$E_{\theta}(k) \sim k^{-3}$ &  $\Pi_{\omega}(k) \sim \text{const.}$\\ & &  $\Pi_{\theta}(k) \sim k^{-2}$\\  \hline
       
        \multirow{7}{*}{Moderate}& For $5\leq k \leq 90$:& \\  &$E_u(k) \sim k^{-2.2}$  & $\Pi_u(k) \sim k^{-0.98}$ (Negative) \\ &$E_{\theta}(k) \sim k^{-1.64}$ & $\Pi_{\omega}(k)\sim \text{const.}$\\ & & $\Pi_{\theta} \sim k^{-0.3}$\\ &For $90\leq k \leq 400$:& \\ &$E_{u}(k) \sim k^{-3}$ & $\Pi_{u}(k)$: weak (positive)\\ &$E_{\theta}(k) \sim k^{-1.64}$ & $\Pi_{\omega}(k)\sim \text{const.}$\\ & & $\Pi_{\theta} \sim k^{-0.3}$\\ \hline
        
            \multirow{4}{*}{Weak}&  & \\ &$E_u(k) \sim k^{-2.5}$ & $\Pi_{u}(k)\sim \text{const.}$ (Negative)\\ &$E_{\theta}(k) \sim k^{-1.6}$ &  $\Pi_{\omega}(k) \sim k^{3/4}$\\ & &  $\Pi_{\theta}(k) \sim k^{-0.13}$\\  \hline

\end{tabular}
\label{table:summary}
\end{center}
\end{table}

For strong stratification, as with numerous previous studies, we observe the emerge of a large scale VSHF. This VSHF is explicitly
identifed as being composed
of internal gravity waves, and further is seen to co-exist with smaller scale turbulence. The turbulent portion of the flow follows some aspects
of the traditional enstrophy cascading regime of pure 2D turbulence. In particular, we find a strong, nearly constant, positive enstophy flux,
zero KE flux and and KE spectrum that scales approximately as $k^{-3}$. But, the PE does not act as a passive scalar. Indeed, it exhibits
an approximate $k^{-3}$ spectrum, and a scale dependent $k^{-2}$ forward flux.

Moderate stratification proves to be very interesting, specifically, there is no VSHF and we observe a modified BO scaling for the KE ---
$E_u(k) \sim k^{-11/5}$ at large scales and
an approximate $k^{-3}$ power-law at small scales. Further,
the nature of the KE flux also changes, with upscale or inverse transfer at large scales and a weak forward transfer at smaller scales. The
PE, on the other hand, always flows downscale and its flux is weakly scale dependent (approximately $k^{-0.3}$). The PE spectrum scales as
$k^{-1.64}$, with no signs of a dual scaling like the KE. But, as the expected change in scaling of the PE spectrum is small, it is possible that higher resolution 
simulations may prove to be useful in this regard.

Weak stratification also differs significantly from pure 2D turbulence. In particular, we actually see a positive scale dependent enstrophy flux
($\sim k^{3/4}$) up to the dissipation scale. In agreement with this form of the enstrophy flux, the KE spectrum scales approximately as $k^{-2.5}$.
The KE flux is robustly negative, and the inverse transfer begins at a comparitively smaller scale than with moderate stratification. The
PE flux, once again, is positive and almost scale independent, and the PE spectrum follows an approximate $k^{-1.6}$ scaling law.

Thus, the nature of 2D stably stratified turbulence under large scale random forcing is dependent on the strength of the ambient stratification.
Despite this diversity, we do observe some universal features. Specifically, the total and potential energy always flow downscale,
which is in agreement with 3D stratified turbulence~\cite{lind}. The KE almost never shows a forward transfer (apart
from the weak downscale transfer at small scales in moderate stratification). In addition,
the zero flux of KE in strong stratification, and its upscale transfer at large scales in moderate and weakly stratified cases is in contrast to the
3D scenario. Finally,
apart from the PE spectrum for weak stratification (and its small scale behaviour for moderate stratification),
the scaling exponents observed very closely match dimensional
expectations when we take into account the scale dependent form of the corresponding flux.

\section*{Acknowledgements}
We thank Anirban Guha for useful discussions.  Our numerical simulations were performed on {\em Chaos} clusters of IIT Kanpur and ``Shaheen II" at KAUST supercomputing laboratory, Saudi Arabia. This work was supported by a research grant (Grant No. SERB/F/3279) from Science and Engineering Research Board, India, computational project  k1052 from KAUST, and the grant PLANEX/PHY/2015239 from Indian Space Research Organisation (ISRO), India. JS would also like to acknowledge support from the IISc ISRO Space Technology Cell project ISTC0352.


\end{document}